\documentclass[conference, 9pt]{IEEEtran}
\IEEEoverridecommandlockouts

\usepackage{cite}
\usepackage{amsmath,amssymb,amsfonts}
\usepackage{algorithmic}
\usepackage{graphicx}
\usepackage{booktabs}
\usepackage{multirow}
\usepackage{hyperref}
\usepackage[table,xcdraw]{xcolor}
\def\BibTeX{{\rm B\kern-.05em{\sc i\kern-.025em b}\kern-.08em
    T\kern-.1667em\lower.7ex\hbox{E}\kern-.125emX}}

\begin{document}

\title{\texttt{voc2vec}: A Foundation Model for Non-Verbal Vocalization}

\author{\IEEEauthorblockN{Alkis Koudounas}
\IEEEauthorblockA{
\textit{Politecnico di Torino}\\
Turin, Italy \\
alkis.koudounas@polito.it}
\and
\IEEEauthorblockN{Moreno La Quatra}
\IEEEauthorblockA{
\textit{Kore University of Enna}\\
Enna, Italy \\
moreno.laquatra@unikore.it
}
\and
\IEEEauthorblockN{Sabato Marco Siniscalchi}
\IEEEauthorblockA{
\textit{Università degli Studi di Palermo}\\
Palermo, Italy \\
sabatomarco.siniscalchi@unipa.it}
\and
\IEEEauthorblockN{Elena Baralis}
\IEEEauthorblockA{
\textit{Politecnico di Torino}\\
Turin, Italy \\
elena.baralis@polito.it}
}

\maketitle

\begin{abstract}

Speech foundation models have demonstrated exceptional capabilities in speech-related tasks. Nevertheless, these models often struggle with non-verbal audio data, such as vocalizations, baby crying, etc., which are critical for various real-world applications.  Audio foundation models well handle non-speech data but also fail to capture the nuanced features of non-verbal human sounds.  In this work, we aim to overcome the above shortcoming and propose a novel foundation model, termed voc2vec, specifically designed for non-verbal human data leveraging exclusively open-soruce non-verbal audio datasets.
We employ a collection of 10 datasets covering around 125 hours of non-verbal audio. 
Experimental results prove that voc2vec is effective in non-verbal vocalization classification, and it outperforms conventional speech and audio foundation models. Moreover, voc2vec consistently outperforms strong baselines, namely OpenSmile and emotion2vec, on six different benchmark datasets. 
To the best of the authors' knowledge, voc2vec is the first universal representation model for vocalization tasks.
\end{abstract}

\begin{IEEEkeywords}
Nonverbal vocalization, Representation Learning, Self-Supervised Models, Pre-trained Models
\end{IEEEkeywords}

\section{Introduction}

Understanding vocal emotional behavior is essential for conversational technologies, such as digital assistants and therapeutic tools, that aim to interact naturally with humans and anticipate their needs~\cite{sorooshconversational, tzirakis2023large, tiantianicassp2024}. Human voice conveys emotion through two primary channels
: speech prosody~\cite{liebenthal2016language} — the melody, rhythm, and timbre of speech — and nonverbal vocalizations, or \textit{``vocal bursts''}~\cite{cowen2019mapping}, such as laughter, sighs, shrieks, and moans. While speech prosody works in tandem with words to communicate emotions, vocal bursts are standalone sounds that express emotion without speech.

Traditionally, research in affective computing has focused on speech prosody for emotional recognition~\cite{tzirakis2018end, eyben2010opensmile}. However, recent studies suggest that vocal bursts may actually convey emotions more directly and powerfully~\cite{cowen2019mapping}. Despite their significance, the study of vocal bursts remains underexplored, largely due to a lack of data capturing a wide variety of these sounds and a lack of models correctly able to understand the nuances of this data. 
Current models, which use datasets designed for speech-based emotion recognition \cite{busso2008iemocap}, have struggled to accurately classify vocal bursts, often focusing only on laughter~\cite{tzirakis2023large, kantharaju2018automatic, condron2021non} and overlooking other meaningful vocalizations like sighs, gasps, or different types of laughs that express amusement, embarrassment, or triumph~\cite{cowen2019mapping}. A recent work~\cite{tzirakis2023large}, fine-tunes transformer-based models on a large-scale and in-the-wild proprietary dataset to detect vocal bursts.

The proliferation of foundation models has significantly advanced the field of natural language processing and speech recognition, enabling remarkable progress in tasks such as speech-to-text transcription, language modeling, and conversational AI.
These models, such as BERT~\cite{devlinbert} and GPT~\cite{gpt} for text, and Wav2Vec~\cite{baevski2020wav2vec}, HuBERT~\cite{hsu2021hubert}, and WavLM~\cite{chen2022wavlm} for speech, have proven highly effective in learning rich and robust representations from large-scale data.
Yet, despite their success in speech-related tasks \cite{superb, italic, ssl_se, koudounas24_interspeech}, these models fall short when applied to non-verbal human sounds, which require a more nuanced understanding of emotional cues~\cite{vaiani2022}. Non-verbal audio — including laughter, crying, and other vocalizations — poses unique challenges that go beyond traditional speech processing. For instance, capturing the subtle acoustic and contextual cues in a baby's cry or the intent behind a sigh demands specialized attention that these models are not optimized to handle. Although audio foundation models like those trained on AudioSet~\cite{audioset} have shown promise in general audio classification tasks, capturing the specific characteristics of non-verbal human sounds remains challenging, leading to suboptimal performance in tasks such as vocalization analysis, and emotion recognition. 
A recent study~\cite{xin2022exploring} suggests using self-supervised learning (SSL) models trained on speech datasets to tackle vocalization tasks. However, the authors employ a non-trivial classifier chain to capture the label dependencies between emotions.

In response to these limitations, we introduce voc2vec, a novel foundation model tailored specifically for non-verbal audio processing. Pre-trained using SSL on a diverse set of 10 open-source non-verbal audio datasets totaling 125 hours, voc2vec is designed to capture the unique features of non-verbal sounds. We explore three pretraining strategies: training the model from scratch, training an SSL model pre-trained on the LibriSpeech dataset, and training an SSL model pre-trained on AudioSet. This targeted approach aims to bridge the gap left by existing speech and audio foundation models, enhancing the model's ability to recognize and interpret emotional vocalizations in diverse human interactions.

Through extensive experiments, we demonstrate that voc2vec significantly outperforms existing models on six benchmark datasets for vocalization tasks, including baby crying classification, emotion recognition, and other non-verbal sound recognition tasks. Our model achieves substantial improvements compared to previous state-of-the-art approaches in key evaluation metrics. Specifically, we register an average 5\% improvement in UAR, 2\% in accuracy, and 4\% in F1 Macro score across the six datasets compared to the next best-performing model, while more than doubling the UAR performance of OpenSmile~\cite{opensmile} and emotion2vec~\cite{ma2023emotion2vec} baselines.

By releasing voc2vec as an open-source model\footnote{Pre-trained models and training scripts are available at \texttt{\url{https://github.com/koudounasalkis/voc2vec}}}, we aim to provide the research community and industry practitioners with a powerful tool for non-verbal audio processing. We believe that voc2vec represents a step forward in the development of foundation models that go beyond speech and enable new applications in areas where understanding non-verbal human sounds is crucial, such as early childhood development and mental health monitoring.

\begin{table}[]
\caption{Statistics of the datasets used for pre-training, including duration (hours), number of samples, average duration.}
\label{table:pretraining_datasets}
\centering
\begin{tabular}{llccc}
\toprule
\multicolumn{2}{l}{\textbf{Dataset}} & \textbf{Dur. (h)} & \textbf{\# Samples} & \textbf{Avg Dur. (s)} \\
\midrule
\multicolumn{2}{l}{AudioSet (vocalization)~\cite{audioset}} 
    & 36.94 
    & 13439 
    & 9.90   \\
\multicolumn{2}{l}{FreeSound (babies)~\cite{freesound}}      
    & 23.42 
    & 1450
    & 58.15  \\
\multicolumn{2}{l}{HumanVoiceDataset}       
    & 0.06  
    & 179 
    & 1.21   \\
\multicolumn{2}{l}{NNIME~\cite{nnime}}                  
    & 3.55  
    & 5596 
    & 2.28   \\
\multicolumn{2}{l}{NonSpeech7K~\cite{nonspeech7k}}             
    & 6.72 
    & 6983 
    & 3.46   \\
\multicolumn{2}{l}{ReCANVo~\cite{recanvo}}                 
    & 2.46 
    & 7077 
    & 1.25   \\
\multicolumn{2}{l}{SingingDatabase~\cite{singingdatabase}}         
    & 3.97 
    & 113 
    & 126.48 \\
\multicolumn{2}{l}{TUT (babies)~\cite{tut}}            
    & 13.17 
    & 1540 
    & 30.79  \\
\multicolumn{2}{l}{VocalSketch~\cite{vocalsketch}}             
    & 10.53 
    & 10705 
    & 3.54 \\
\multicolumn{2}{l}{VocalSound~\cite{vocalsound}}              
    & 24.37 
    & 20985 
    & 4.18   \\ \midrule
\multicolumn{2}{l}{\textbf{Voc125 (Total)}}          
    & \textbf{125.19} 
    & \textbf{68067} 
    & \textbf{6.67} \\
\bottomrule
\end{tabular}
\end{table}

\section{Methodology}

Voc2vec is built upon the wav2vec 2.0 framework~\cite{baevski2020wav2vec}, which has proven effective for self-supervised learning from audio. 
The model consists of two main components: a multi-layer convolutional neural network (CNN) encoder and a Transformer network.
The CNN encoder extracts lower-dimensional latent representations from raw audio input, while the Transformer captures contextual relationships over time, allowing the model to process both local and global vocalization features.
Given a raw audio input $X = \{x_1, x_2, \dots, x_T\}$, the CNN encoder outputs a sequence of latent representations:
\begin{equation}
    \mathbf{Z} = \text{Encoder}(X) = \{z_1, z_2, \dots, z_T'\}, \quad T' < T,
\end{equation}
where $T'$ is the number of frames in the encoded sequence, and each $z_t \in \mathbb{R}^d$ is a $d$-dimensional vector capturing the features of the audio input of each frame. 
The Transformer operates on these representations, extracting contextualized vectors $\mathbf{C} = \{c_1, c_2, \dots, c_T'\}$ that incorporate local acoustic features as well as broader temporal dependencies.
This architecture enables voc2vec to learn rich and robust representations from raw audio data, which are essential for capturing the variety of non-verbal vocalizations.

\subsection{Pre-training Datasets}

The core contribution of voc2vec lies in the careful selection of diverse, open-source datasets for pre-training, specifically chosen to focus on non-verbal vocalizations. 
These datasets collectively cover around 125 hours of audio, ensuring that the model is exposed to a wide variety of human vocalizations, typically underrepresented in speech datasets.
Each dataset is chosen to represent different forms of non-verbal communication, such as emotional bursts, human reactions, and environmental sounds that involve vocal interaction. 
The datasets used for pre-training are summarized in Table~\ref{table:pretraining_datasets}.

\vspace{1mm}
\noindent
\textit{AudioSet (vocalization split) \cite{audioset}} contains classes such as \textit{laughter}, \textit{crying \& sobbing}, \textit{chewing}, among others. 
Selected categories focus on non-verbal vocalizations in an ``in-the-wild'' setting, providing a diverse range of types and contexts.

\vspace{1mm}
\noindent
\textit{FreeSound (babies split) \cite{freesound}} is a rich collection of infant vocalizations, including crying, laughing, and babbling. 
It is particularly useful for tasks like baby cry detection, which require recognizing emotional content in infant vocal cues.

\vspace{1mm}
\noindent
\textit{Human Voice Dataset\footnote{\texttt{\url{https://github.com/vocobox/human-voice-dataset}}}} offers a small but unique set of recordings focused on singing techniques, including pitch control and 
production of vowels and consonants. 
This dataset contributes to modeling non-speech audio by exposing voc2vec to controlled vocal expressions. 

\vspace{1mm}
\noindent
\textit{NNIME \cite{nnime}} is a Chinese affective dataset that features dyadic interactions in naturalistic settings. 
Each interaction captures affective states such as anger, happiness, sadness, and frustration, helping voc2vec learn how non-verbal vocalizations express emotions in real-world conversational scenarios.

\vspace{1mm}
\noindent
\textit{NonSpeech7K \cite{nonspeech7k}} includes a broad range of non-speech human sounds, such as breathing, coughing, sneezing, laughing, and screaming. 
It enriches voc2vec with non-verbal audio samples that may frequently occur in daily life, enhancing model generalization across different vocalization types.

\vspace{1mm}
\noindent
\textit{ReCANVo \cite{recanvo}} focuses on communicative and affective vocalizations from minimally speaking individuals. 
It includes over 7000 vocalizations, making it valuable for understanding how non-verbal sounds communicate emotions and needs.

\vspace{1mm}
\noindent
\textit{Singing Database \cite{singingdatabase}} contains over 70 recordings from singers, primarily focused on Chinese Opera. 
This dataset introduces voc2vec to structured and controlled vocal techniques, allowing the model to differentiate between singing and other non-verbal sounds.

\vspace{1mm}
\noindent
\textit{TUT (babies split) \cite{tut}} features baby cries recorded in environments with background noise. 
This dataset helps voc2vec to learn how to detect infant vocalizations in noisy settings, which is critical for real-world applications.

\vspace{1mm}
\noindent
\textit{VocalSketch \cite{vocalsketch}} is an uncommon dataset of vocal imitations of environmental sounds. 
Participants were asked to imitate various sounds using their voices, which provides voc2vec with a unique set of non-verbal sounds that mimic real-world noises, expanding the model's versatility.

\vspace{1mm}
\noindent
\textit{VocalSound \cite{vocalsound}} is a crowdsourced dataset containing over 21,000 recordings of laughter, sighs, sneezes, coughs, and other spontaneous non-verbal expressions. 
This dataset is essential for voc2vec’s ability to generalize across spontaneous and often subtle vocal bursts encountered in daily life.

\vspace{1mm}
Together, these datasets aims at providing voc2vec with a diverse and comprehensive foundation of non-verbal sounds, despite the relatively smaller total duration compared to larger speech collections.

\section{Experimental Setup}

\subsection{Pre-training Procedure}
Our voc2vec model follows the wav2vec 2.0 pre-training setup, adapted for non-verbal audio processing. 
The feature encoder consists of seven convolutional blocks with 512 channels, using strides of (5, 2, 2, 2, 2, 2, 2) and kernel widths of (10, 3, 3, 3, 3, 2, 2). This setup provides an encoder output frequency of 49 Hz with a stride of approximately 20ms between samples and a receptive field covering 25ms of audio.
We adopt the same architecture as the base configuration in wav2vec 2.0, which contains 12 transformer blocks, a model dimension of 768, an inner dimension of 3,072, and 8 attention heads. The pre-training is done on two A6000 GPUs over 10.6 days.
We use the Adam optimizer, warming up the learning rate to a peak of 5e-4 during the first 8\% of updates, followed by linear decay. 
The total pre-training consists of 400k updates.
All pre-training datasets are pre-processed to include audio samples of 10 seconds in length for consistent model input during pre-training.
The best checkpoint, during pre-training, is selected according to the lowest validation loss.


We explore three different pre-training strategies to evaluate how various initializations impact voc2vec’s performance. 
The first strategy pre-trains the model from scratch using only the non-verbal vocalization datasets (\textsc{Voc125}), focusing exclusively on vocalization data. 
The second strategy continues pre-training from a model that was initially trained on the LibriSpeech dataset~\cite{baevski2020wav2vec}, while the third strategy builds upon a model pre-trained on AudioSet~\cite{arch}.
These strategies allow us to assess whether openly-available non-verbal data can complement or enhance pre-existing models trained on larger speech or general audio datasets. 

\subsection{Fine-Tuning}
Following pre-training, we fine-tune voc2vec on labeled datasets by adding a randomly initialized output layer on top of the Transformer to adapt the model to specific downstream tasks. 
In all cases, the model is trained with a consistent setup: a batch size of 16 for up to 50 epochs, with early stopping applied if no improvement is observed over 5 consecutive epochs.
The learning rate is linearly increased to $10^{-4}$ over the first 500 warmup steps, and then linearly decayed for the remaining fine-tuning steps.
As the datasets do not provide predefined train-validation-test splits, we implement 10-fold cross-validation for a robust and reliable evaluation. 
This configuration is maintained across all tested models to ensure fair comparison.

\subsection{Fine-tuning Datasets}
We evaluate voc2vec on six classification tasks using diverse datasets, each covering different types of non-verbal vocalizations. 
The datasets and their characteristics are summarized in Table~\ref{table:finetuning_datasets}. 
First, we use ASVP-ESD~\cite{asvp}, which contains 12,625 audio files of both speech and non-speech emotional sounds, collected from real-world sources like movies and YouTube, with an additional 1,204 baby vocalizations. 
We test voc2vec on both the full dataset and the baby-specific subset (e.g., ASPV-ESD (babies)).
For infant-specific tasks, we also use the Donate a Cry dataset\footnote{\texttt{\url{https://github.com/gveres/donateacry-corpus}}}, which consists of around 1000 baby cry sound clips categorized into actions (e.g., burping) or emotional states (e.g., discomfort).
The CNVVE dataset~\cite{cnvve} encompasses 950 audio samples across 6 categories of vocal expressions, collected from 42 participants, both healthy and dysarthric. It serves as a significant benchmark for non-verbal voice recognition in human-computer interaction and assistive technology.
The NonVerbal Vocalization Dataset\footnote{\texttt{\url{https://www.openslr.org/99/}}} includes a variety of non-verbal sounds such as \textit{laughing} and \textit{crying}, making it an ideal candidate for evaluating the model in general non-verbal sound classification tasks.
Finally, the VIVAE dataset~\cite{vivae} contains 1,085 human vocalizations categorized in 6 classes, representing both positive and negative emotional states across varying levels of intensity. 
\begin{table}[]
\caption{Statistics of fine-tuning datasets, including number of classes, duration (h), number of samples and avg duration.}
\label{table:finetuning_datasets}
\centering
\scalebox{0.88}{%
\begin{tabular}{llcccc}
\toprule
\multicolumn{2}{l}{\textbf{Dataset}} & \multicolumn{1}{l}{\textbf{\# Classes}} & \textbf{Dur. (h)} & \textbf{\# Samples} & \textbf{\# Avg Dur. (s)} \\
\midrule
\multicolumn{2}{l}{ASVP-ESD~\cite{asvp}}
    & 13 
    & 15.07 
    & 12625 
    & 4.30 \\
\multicolumn{2}{l}{ASVP-ESD (babies)~\cite{asvp}}
    & 7  
    & 2.91  
    & 1339  
    & 8.22 \\
\multicolumn{2}{l}{CNVVE~\cite{cnvve}}     
    & 6  
    & 0.2   
    & 921   
    & 0.78 \\
\multicolumn{2}{l}{Donate A Cry}
    & 5  
    & 0.88  
    & 457   
    & 6.93 \\
\multicolumn{2}{l}{NonVerbal Vocalization} 
    & 16 
    & 0.6   
    & 800   
    & 3.10 \\
\multicolumn{2}{l}{VIVAE~\cite{vivae}}
    & 6  
    & 0.27  
    & 1085  
    & 0.90 \\
\bottomrule
\end{tabular}}
\end{table}

\begin{table}[]
\caption{Ablation study on the initialization strategies of voc2vec. 
Mean $\pm$ std for UAR, Accuracy and F1 Macro across the six downstream datasets. For comparison, we also include wav2vec 2.0 and HuBERT models pre-trained on LS.}
\label{table:ablation-initialization}
\centering
\scalebox{0.91}{%
\begin{tabular}{lcccc}
\toprule
\multicolumn{1}{c}{\textbf{Model}} & \textbf{Pre-training DS} & \textbf{UAR} & \textbf{Accuracy} & \textbf{F1 Macro} \\
\midrule
\texttt{voc2vec} 
    & Voc125 
    & .612\scriptsize±.212 
    & .729\scriptsize±.146 
    & .580\scriptsize±.230 \\
\texttt{voc2vec-as} 
    & AS+Voc125 
    & .603\scriptsize±.183 
    & .754\scriptsize±.131 
    & .574\scriptsize±.194 \\
\texttt{voc2vec-ls} 
    & LS960+Voc125 
    & \cellcolor[HTML]{D9EAD3}\textbf{.661\scriptsize±.206} 
    & \cellcolor[HTML]{D9EAD3}\textbf{.802\scriptsize±.139} 
    & \cellcolor[HTML]{D9EAD3}\textbf{.636\scriptsize±.223} \\
\midrule
\texttt{wav2vec2-ls} 
    & LS960
    & .599\scriptsize±.237
    & .739\scriptsize±.192 
    & .569\scriptsize±.259 \\
\texttt{hubert-ls} 
    & LS960
    & .627\scriptsize±.214 
    & .784\scriptsize±.149 
    & .611\scriptsize±.222 \\
\bottomrule
\end{tabular}}
\end{table}

\begin{table*}[]
\caption{10-Fold CV fine-tuning results across the explored six datasets.}
\vspace{-1mm}
\label{table:results}
\centering
\begin{tabular}{l|ccc|ccc|ccc}
\toprule
& \textbf{UAR}
    & \textbf{Accuracy} 
    & \textbf{F1 Macro} 
    & \textbf{UAR} 
    & \textbf{Accuracy} 
    & \textbf{F1 Macro} 
    & \textbf{UAR} 
    & \textbf{Accuracy} 
    & \textbf{F1 Macro} \\
\cmidrule{2-10}
\multirow{-2}{*}{\textbf{Model}} 
    & \multicolumn{3}{c}{\textbf{ASPV-ESD}} 
    & \multicolumn{3}{c}{\textbf{ASPV-ESD (babies)}} 
    & \multicolumn{3}{c}{\textbf{CNVVE}} \\
\midrule
\texttt{wav2vec2-as}
    & .590\scriptsize±.016 
    & .624\scriptsize±.014 
    & .577\scriptsize±.016 
    & .521\scriptsize±.126 
    & .890\scriptsize±.044 
    & .460\scriptsize±.132 
    & .839\scriptsize±.060 
    & .838\scriptsize±.063 
    & .809\scriptsize±.073 \\
\texttt{wav2vec2-ls}
    & .626\scriptsize±.014 
    & .672\scriptsize±.013 
    & \cellcolor[HTML]{D9EAD3}\textbf{.627\scriptsize±.013} 
    & .432\scriptsize±.171 
    & .891\scriptsize±.059 
    & .378\scriptsize±.140 
    & .971\scriptsize±.018 
    & .970\scriptsize±.017
    & .973\scriptsize±.016 \\
\texttt{hubert-as}
    & .587\scriptsize±.173 
    & .621\scriptsize±.194 
    & .577\scriptsize±.183 
    & .456\scriptsize±.121 
    & .856\scriptsize±.055 
    & .389\scriptsize±.106 
    & .919\scriptsize±.058 
    & .922\scriptsize±.058 
    & .918\scriptsize±.059 \\
\texttt{hubert-ls}
    & .622\scriptsize±.017 
    & .664\scriptsize±.011 
    & .619\scriptsize±.014 
    & .515\scriptsize±.134 
    & .924\scriptsize±.023 
    & \cellcolor[HTML]{D9EAD3}\textbf{.505\scriptsize±.141} 
    & .972\scriptsize±.017 
    & .972\scriptsize±.017 
    & .972\scriptsize±.017 \\
\texttt{wavlm}
    & .615\scriptsize±.016 
    & .663\scriptsize±.012 
    & .609\scriptsize±.016 
    & .501\scriptsize±.151 
    & .870\scriptsize±.063
    & .434\scriptsize±.144 
    & .976\scriptsize±.018 
    & .975\scriptsize±.019 
    & .975\scriptsize±.019 \\
\texttt{wavlm-plus} 
    & .594\scriptsize±.017 
    & .649\scriptsize±.009 
    & .591\scriptsize±.016 
    & .441\scriptsize±.133 
    & .860\scriptsize±.080 
    & .385\scriptsize±.155 
    & .978\scriptsize±.015 
    & .978\scriptsize±.015 
    & .978\scriptsize±.015 \\
\texttt{data2vec} 
    & .585\scriptsize±.029 
    & .606\scriptsize±.027 
    & .567\scriptsize±.030 
    & .514\scriptsize±.171 
    & .880\scriptsize±.043 
    & .468\scriptsize±.165 
    & .981\scriptsize±.008 
    & .980\scriptsize±.007 
    & .981\scriptsize±.008 \\
\texttt{emotion2vec} 
    & .575\scriptsize±.042 
    & .618\scriptsize±.012 
    & .537\scriptsize±.018 
    & .509\scriptsize±.206 
    & \cellcolor[HTML]{D9EAD3}\textbf{.926\scriptsize±.330} 
    & .467\scriptsize±.175 
    & .955\scriptsize±.022 
    & .950\scriptsize±.023 
    & .951\scriptsize±.023 \\
\texttt{opensmile} 
    & .106\scriptsize±.055 
    & .125\scriptsize±.037 
    & .040\scriptsize±.011 
    & .281\scriptsize±.137 
    & .785\scriptsize±.192 
    & .258\scriptsize±.126 
    & .206\scriptsize±.092 
    & .249\scriptsize±.053 
    & .160\scriptsize±.050 \\
\texttt{voc2vec-ls} 
    & \cellcolor[HTML]{D9EAD3}\textbf{.633\scriptsize±.009} 
    & \cellcolor[HTML]{D9EAD3}\textbf{.673\scriptsize±.009} 
    & \cellcolor[HTML]{D9EAD3}\textbf{.627\scriptsize±.006} 
    & \cellcolor[HTML]{D9EAD3}\textbf{.526\scriptsize±.189} 
    & .914\scriptsize±.039 
    & .491\scriptsize±.175
    & \cellcolor[HTML]{D9EAD3}\textbf{.982\scriptsize±.011} 
    & \cellcolor[HTML]{D9EAD3}\textbf{.982\scriptsize±.010} 
    & \cellcolor[HTML]{D9EAD3}\textbf{.982\scriptsize±.010} \\
\midrule
\midrule
\textbf{Model} 
    & \multicolumn{3}{c}{\textbf{NonVerbal Vocalization}} 
    & \multicolumn{3}{c}{\textbf{Donate a Cry}} 
    & \multicolumn{3}{c}{\textbf{VIVAE}} \\
\midrule
\texttt{wav2vec2-as} 
    & .619\scriptsize±.065 
    & .607\scriptsize±.070 
    & .575\scriptsize±.064 
    & .282\scriptsize±.086 
    & .783\scriptsize±.065 
    & .254\scriptsize±.078 
    & .449\scriptsize±.058 
    & .448\scriptsize±.059 
    & .412\scriptsize±.059 \\
\texttt{wav2vec2-ls} 
    & .838\scriptsize±.067 
    & .817\scriptsize±.088 
    & .818\scriptsize±.079 
    & .360\scriptsize±.071 
    & .709\scriptsize±.153 
    & .295\scriptsize±.075 
    & .365\scriptsize±.162 
    & .375\scriptsize±.158 
    & .320\scriptsize±.194 \\
\texttt{hubert-as} 
    & .837\scriptsize±.048 
    & .826\scriptsize±.051 
    & .803\scriptsize±.048 
    & .364\scriptsize±.075 
    & .743\scriptsize±.111 
    & .304\scriptsize±.064 
    & .455\scriptsize±.059 
    & .464\scriptsize±.055 
    & .415\scriptsize±.065 \\
\texttt{hubert-ls} 
    & .812\scriptsize±.047 
    & .810\scriptsize±.047 
    & .789\scriptsize±.054 
    & .312\scriptsize±.067 
    & .802\scriptsize±.051 
    & .276\scriptsize±.062 
    & .527\scriptsize±.067 
    & .534\scriptsize±.062 
    & .510\scriptsize±.071 \\
\texttt{wavlm}
    & .840\scriptsize±.050
    & .832\scriptsize±.041 
    & .820\scriptsize±.054 
    & .352\scriptsize±.093 
    & .721\scriptsize±.101 
    & .307\scriptsize±.058 
    & .230\scriptsize±.061 
    & .256\scriptsize±.074 
    & .161\scriptsize±.067 \\
\texttt{wavlm-plus} 
    & .731\scriptsize±.035
    & .728\scriptsize±.046 
    & .676\scriptsize±.043 
    & .349\scriptsize±.096 
    & .835\scriptsize±.041 
    & .306\scriptsize±.073 
    & .194\scriptsize±.056 
    & .211\scriptsize±.085 
    & .107\scriptsize±.073 \\
\texttt{data2vec} 
    & .707\scriptsize±.076 
    & .706\scriptsize±.078 
    & .651\scriptsize±.073 
    & .306\scriptsize±.106 
    & .768\scriptsize±.100 
    & .252\scriptsize±.081 
    & .301\scriptsize±.096 
    & .307\scriptsize±.101 
    & .224\scriptsize±.117 \\
\texttt{emotion2vec} 
    & .764\scriptsize±.038 
    & .760\scriptsize±.035
    & .726\scriptsize±.046 
    & .210\scriptsize±.051 
    & \cellcolor[HTML]{D9EAD3}\textbf{.837\scriptsize±.036} 
    & .228\scriptsize±.054 
    & .513\scriptsize±.082 
    & .516\scriptsize±.076 
    & .502\scriptsize±.075 \\
\texttt{opensmile} 
    & .021\scriptsize±.007 
    & .094\scriptsize±.030 
    & .026\scriptsize±.015 
    & .212\scriptsize±.050 
    & .784\scriptsize±.155 
    & .222\scriptsize±.062 
    & .128\scriptsize±.077 
    & .128\scriptsize±.025 
    & .092\scriptsize±.025 \\
\texttt{voc2vec-ls} 
    & \cellcolor[HTML]{D9EAD3}\textbf{.872\scriptsize±.049} 
    & \cellcolor[HTML]{D9EAD3}\textbf{.869\scriptsize±.036} 
    & \cellcolor[HTML]{D9EAD3}\textbf{.848\scriptsize±.049} 
    & \cellcolor[HTML]{D9EAD3}\textbf{.378\scriptsize±.173} 
    & .793\scriptsize±.095 
    & \cellcolor[HTML]{D9EAD3}\textbf{.311\scriptsize±.092} 
    & \cellcolor[HTML]{D9EAD3}\textbf{.573\scriptsize±.062} 
    & \cellcolor[HTML]{D9EAD3}\textbf{.578\scriptsize±.055} 
    & \cellcolor[HTML]{D9EAD3}\textbf{.558\scriptsize±.060} \\
\bottomrule
\end{tabular}
\end{table*}

\begin{figure*}[!ht]
  \centering
  \includegraphics[width=\linewidth]{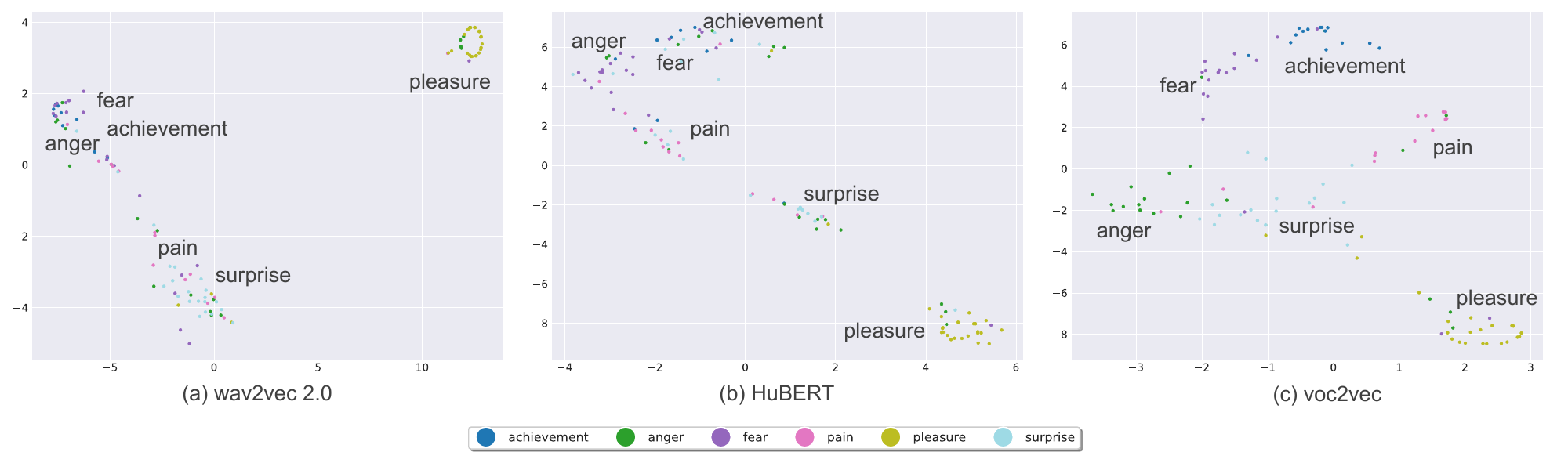}
  \caption{\textbf{t-SNE visualization.} \textsc{VIVAE} dataset, first fold. \texttt{wav2vec-ls} (left), \texttt{hubert-ls} (center), \texttt{voc2vec-ls} (right).}
    \vspace{-3mm}
  \label{fig:tsne}
\end{figure*}

\subsection{Baselines}
We evaluate voc2vec in its three variants, all pre-trained using SSL on \textsc{Voc125}: cold-start \texttt{voc2vec}, i.e., trained from scratch, \texttt{voc2vec-ls} initialized from LibriSpeech pre-training, and \texttt{voc2vec-as} initialized from AudioSet pre-training. 
Our comparison includes several baseline models based on wav2vec 2.0, HuBERT, WavLM, and data2vec, all of which have demonstrated strong performance in speech-related tasks. 
Specifically, we compare voc2vec with wav2vec 2.0 pre-trained on LibriSpeech~\cite{baevski2020wav2vec} (\texttt{wav2vec2-ls}) and AudioSet (\texttt{wav2vec2-as}), as well as HuBERT pre-trained on LibriSpeech (\texttt{hubert-ls}) and AudioSet (\texttt{hubert-as})~\cite{hsu2021hubert}. 
We also evaluate WavLM~\cite{chen2022wavlm}, WavLM-plus, and data2vec~\cite{baevski2022data2vec}, all in their base configurations, with approximately 90 million parameters.

Additionally, we include two feature-based baselines: OpenSmile~\cite{opensmile} and emotion2vec~\cite{ma2023emotion2vec}. 
OpenSmile is a widely used feature extraction toolkit, while emotion2vec is a more recent model that achieves state-of-the-art results in speech emotion recognition. 
For both baselines, we extract features and pass them through two linear layers with ReLU activation for a comparable evaluation setup.

\subsection{Evaluation Metrics}

To assess the performance of voc2vec and the baseline models, we employ a set of complementary metrics that capture different aspects of classification performance. 

\vspace{1mm}
\noindent
\textit{Unweighted Average Recall (UAR)} measures the average recall across all classes, giving equal weight to each class regardless of its frequency. This is particularly useful for imbalanced datasets, ensuring that performance on minority classes is fairly represented.

\vspace{1mm}
\noindent
\textit{Accuracy} is the ratio of correctly predicted instances to the total number of instances. While accuracy gives an overall performance measure, it may be skewed on imbalanced datasets, favoring the majority class.

\vspace{1mm}
\noindent
\textit{F1 Macro Score} is the harmonic mean of precision and recall, calculated separately for each class and then averaged. It gives equal weight to each class and is effective in handling class imbalances.


\section{Results and Discussion}


To first evaluate the effect of different pre-training strategies on voc2vec, we conducted an ablation study, presented in Table~\ref{table:ablation-initialization}, comparing three variants: \texttt{voc2vec} trained from scratch, \texttt{voc2vec-as} initialized from AudioSet pre-training, and \texttt{voc2vec-ls} initialized from LibriSpeech pre-training. 
The results, showing the average performance metrics on the six downstream datasets, clearly demonstrate that \texttt{voc2vec-ls} performs best. Starting from a model pre-trained on LibriSpeech thus provides a strong foundation for non-verbal vocalization tasks. 
In comparison, \texttt{voc2vec} trained from scratch shows lower performance across all metrics, highlighting the challenges of training on limited data. 
The AudioSet-initialized model, \texttt{voc2vec-as}, shows only a slight improvement over \texttt{voc2vec}, indicating that pre-training on speech data offers a more beneficial starting point than general audio datasets like AudioSet.

When compared with models pre-trained solely on LibriSpeech (\texttt{wav2vec2-ls} and \texttt{hubert-ls}), \texttt{voc2vec-ls} shows clear improvements. 
It outperforms \texttt{wav2vec2-ls} by 10.3\% in UAR and 11.7\% in F1 Macro, emphasizing the benefit of further pre-training on domain-specific data after initializing from a robust speech model.

Table~\ref{table:results} presents our benchmarking results. voc2vec consistently achieves state-of-the-art performance across the majority of datasets and evaluation metrics. 
We focus here on the best-performing variant, \texttt{voc2vec-ls}, which consistently outperforms both baseline models and other voc2vec variants, as previously shown. 
\texttt{voc2vec-ls} demonstrates clear superiority across all datasets, excelling in the NonVerbal Vocalization and VIVAE datasets, where it achieves the highest UAR, Accuracy, and F1 Macro scores.
For the former, \texttt{voc2vec-ls} achieves a UAR of 0.872, significantly outperforming the closest competitor, \texttt{wavlm}, by 3.8\% in UAR. 
In the VIVAE dataset, \texttt{voc2vec-ls} maintains a strong lead, with a UAR of 0.573 and F1 Macro of 0.558, surpassing all other models, including \texttt{hubert-ls}, which ranks second. 
Compared to baseline models like \texttt{wav2vec2-as} and \texttt{hubert-as}, \texttt{voc2vec-ls} consistently performs better across all tasks. 
Although \texttt{emotion2vec} performs well on the Donate a Cry dataset with an Accuracy of 0.837, \texttt{voc2vec-ls} remains highly competitive, leading in F1 Macro and Accuracy, which underscores its balanced performance across classes.

Figure~\ref{fig:tsne} finally plots t-SNE projections of the embeddings from \texttt{wav2vec2-ls} (a), \texttt{hubert-ls} (b), and \texttt{voc2vec-ls} (c) for the VIVAE dataset. Although not flawless, voc2vec representations show a much more consistent and cohesive clustering compared to those from wav2vec 2.0 and HuBERT. Our model demonstrates improved intra-cluster cohesion and inter-cluster separation.




\section{Conclusion}
In this paper, we introduced voc2vec, a foundation model specifically designed for non-verbal audio processing, trained using self-supervised learning on 10 diverse non-verbal audio datasets. Our model significantly outperforms existing SSL models across multiple benchmark vocalization tasks, with notable improvements in UAR, accuracy, and F1 Macro score. By releasing voc2vec as an open-source tool, we aim to advance the understanding and classification of non-verbal human sounds.


\bibliographystyle{IEEEtran}
\bibliography{main}

\end{document}